\documentclass[journal]{IEEEtran}
\IEEEoverridecommandlockouts
\usepackage[utf8]{inputenc}
\usepackage[english]{babel}
\usepackage{cite}
\usepackage{amsmath,amssymb,amsfonts}
\usepackage{graphicx}
\usepackage{textcomp}
\usepackage{xcolor}
\usepackage{bm,cite,amsmath,amssymb,amsthm}
\def\BibTeX{{\rm B\kern-.05em{\sc i\kern-.025em b}\kern-.08em
    T\kern-.1667em\lower.7ex\hbox{E}\kern-.125emX}}
\usepackage{cuted}
\usepackage{flushend}
\usepackage{algorithm}    
\usepackage[noend]{algpseudocode} 
\usepackage{amsmath}
\usepackage{comment}

\begin{document}

\title{QoE Optimization for Live Video Streaming  in UAV-to-UAV
Communications via Deep Reinforcement Learning
}
\author{Liyana~Adilla~binti~Burhanuddin,~\IEEEmembership{Student,~IEEE,}        Xiaonan~Liu,~\IEEEmembership{Student,~IEEE,} Yansha~Deng,~\IEEEmembership{Member,~IEEE,} Ursula~Challita,~\IEEEmembership{Member,~IEEE,}        and~András~Zahemszky,~\IEEEmembership{Member,~IEEE}  
\thanks{L. A. B. Burhanuddin, X. Liu and Y.  Deng are with Department of Engineering, King's College London, London, UK. L. A. B. Burhanuddin is also with School of Information Science and Technology, Xiamen University Malaysia, Sepang, Malaysia (e-mail: liyana.burhanuddin@kcl.ac.uk, xiaonan.liu@kcl.ac.uk, yansha.deng@kcl.ac.uk).}
\thanks{U. Challita and A. Zahemszky are with Ericsson AB, Stockholm, Sweden (email: ursula.challita@ericsson.com, andras.zahemszky@ericsson.com)

\textit{Corresponding author: Yansha Deng} } 
}

\maketitle

\begin{abstract}
A challenge for rescue teams when fighting against wildfire in remote areas is the lack of information, such as the size and images of fire areas. As such, live streaming from Unmanned Aerial Vehicles (UAVs), capturing videos of dynamic fire areas, is crucial for firefighter commanders in any location to monitor the fire situation with quick response. The 5G network is a promising wireless technology to support such scenarios. In this paper, we consider a UAV-to-UAV (U2U) communication scenario, where a UAV at a  high altitude acts as a mobile base station (UAV-BS) to stream videos from other flying UAV-users (UAV-UEs) through the uplink. Due to the mobility of the UAV-BS and UAV-UEs, it is important to determine the optimal movements and transmission powers for UAV-BSs and UAV-UEs in real-time,  so as to maximize the data rate of video transmission with smoothness and low latency, while mitigating the interference according to the dynamics in fire areas and wireless channel conditions. In this paper, we co-design the video resolution, the movement, and the power control of UAV-BS and UAV-UEs  to maximize the Quality of Experience (QoE) of real-time video streaming. To learn the dynamic fire areas and communication environment, we apply the Deep Q-Network (DQN) and Actor-Critic (AC) to maximize the QoE of video transmission from all UAV-UEs to a single UAV-BS. Simulation results show the effectiveness of our proposed algorithm in terms of the QoE, delay and video smoothness as compared to the Greedy algorithm.

\end{abstract}


\section{Introduction}
Over the years, an increasing number of wildfires has caused challenges for firefighters to control and monitor fire in remote areas \cite{muller2020forest,sung2019primo_firefighting}. Without new technology to monitor the incident area from the control station, the current practice of the fire station control lacks the technology to remotely visualize the dynamic fire situation in real-time for immediate action \cite{sung2019primo_firefighting}. Therefore, monitoring multiple firefighting areas in different locations with dynamic fire heights and areas is vital. Unmanned Aerial Vehicles (UAVs) with low cost, high mobility, and the capability to capture high-definition video, can be a good solution to oversee the fire situation, and facilitate the fire commander’s response for the  choice of number of firefighters and firefighting machines. The use of UAVs provides the fire commander with the overall situation of the fire and danger, such as explosions or human requiring rescue. More importantly, it helps to reduce any imminent dangers and obstacles to firefighters. Existing wireless technologies, such as WiFi, Bluetooth, and radio wave, can only support UAVs’ communication within a short transmission range, which are inefficient for multi-UAV collaboration with limited multi-UAV control \cite{azari2019u2u}. Meanwhile, cellular networks can support the real-time video streaming from UAV  users (UAV-UEs) with beyond line of sight control, low latency, real-time communication, and ubiquitous coverage from flying base stations (UAV-BSs) with wireless backhaul to the core networks.
Despite the growing interests in cellular-connected UAVs, there are still many challenges unsolved for commercial deployment \cite{azari2019u2u} \cite{zhang2019u2x}. A UAV has been initially proposed as a relay to help other UAVs transmit to a nearby terrestrial base station (BS) with low signal to noise ratio (SNR)   \cite{zhang2019u2x}. 
When the distance of UAV-to-UAV (U2U) communication decreases, the SNR of the transmission among the UAVs increases resulting in a better transmission performance\cite{azari2020uav}.

The use of UAVs in disaster scenarios has been investigated in literature   \cite{UAV_disaster,joshi2020simulation,masaracchia2020concept,challita2018deep,sadi2014minimum,zhang2017joint,padilla2020flight,selim2019outage}. In \cite{UAV_disaster}, the UAV was introduced as an emergency BS to serve the affected ground users with limited coverage. In \cite{joshi2020simulation}, multiple mini-UAVs were used to form flying ad-hoc network (FANET) to explore large and disjoint terrain in disaster areas while adapting their transmission power to optimize the energy usage.  In \cite {zhang2017joint}, through optimizing the trajectory and the transmit power of the UAV and the mobile device, the outage probability of the UAV relay network in the disaster area was minimized. In \cite {padilla2020flight}, a UAV platform was developed to compensate the communication loss during a natural disaster, with the aim to obtain the optimal flight paths in high-rise urban and urban microcell environment. In \cite{selim2019outage}, UAV-assisted networks was studied in disaster area, and the proposed power control optimization problem was solved via relaxing the non-convex problem.
Nevertheless, no studies have focused on the  real-time video streaming between UAV-UEs and UAV-BS. 

Real-time video streaming has higher requirements in terms of data rate, latency, and smoothness compared to other data types. In a firefighting scenario, the network channel capacity fluctuates dramatically with the dynamic environment alongside the UAVs' movement, which can cause poor network performance and undesirable delays. This in turn makes it harder to learn the pattern variance of the channel capacity, thus resulting in failure to transmit with high capacity and high video quality. To overcome these limitations, the authors in \cite{xiao2019bitrate} applied the Additive Variation Bitrate (ABR) method with Deep Reinforcement Learning (DRL) to select proper video resolution based on previous communication rate and throughput. However, \cite{xiao2019bitrate} only focused on a single video source ABR, which was guided by RL to make decisions based on network observations and video playback states for selecting  the optimal  video resolution. In search and rescue firefighting scenario, a nonordinary optical camera \cite{govil2020preliminary_fire}	should be considered to ensure the reception of a high quality video. To deal with a more complex environment and practical scenarios, such as search and rescue firefighting scenarios, the DRL algorithm is a promising tool for solving the problem of jointly optimizing the UAVs location while maximizing  the data rate \cite{nanjiang2018deep}.

In this paper, 
we consider a cellular-connected UAV-BS
streaming the real-time video captured by UAV-UEs from the
firefighting area for fire monitoring. The contributions of this paper are summarized as follows:
\begin{itemize}
    \item We develop a framework for a dynamic UAV-to-UAV (U2U) communication model with a moving UAV-BS in multiple firefighting areas to capture a live-streaming panoramic view. We model the dynamic fire arrival with different heights in every fire area and UAVs' request arrival as Poisson process in each time slot, and design the UAV-UEs location spaces to capture a full panoramic view with multiple UAVs.

    \item To guarantee the smoothness and latency of the live video streaming among UAV-BS and UAV-UEs in this U2U network, we formulate a long-term Quality of Experience (QoE) maximization problem via optimizing the UAVs' positions, video resolution, and transmit power over each time slot.  
    
    \item To solve the above problem, we propose a Deep Reinforcement Learning (DRL) approach based on the Actor-Critic (AC) and the Deep Q Network (DQN). Our results shown that our proposed AC and DQN approaches outperform the greedy algorithm in terms of QoE. 
\end{itemize}

The rest of this paper is organized as follows. The system model and problem formulation are given in Section \ref{sectio}. The optimization problem via reinforcement learning is presented in Section \ref{method}. Simulation results and conclusion are presented in Sections \ref{sim} and \ref{conclusion}, respectively. 

\begin{figure}[!t]
    \centering
    \includegraphics[width=3.0in,height=2.3in]{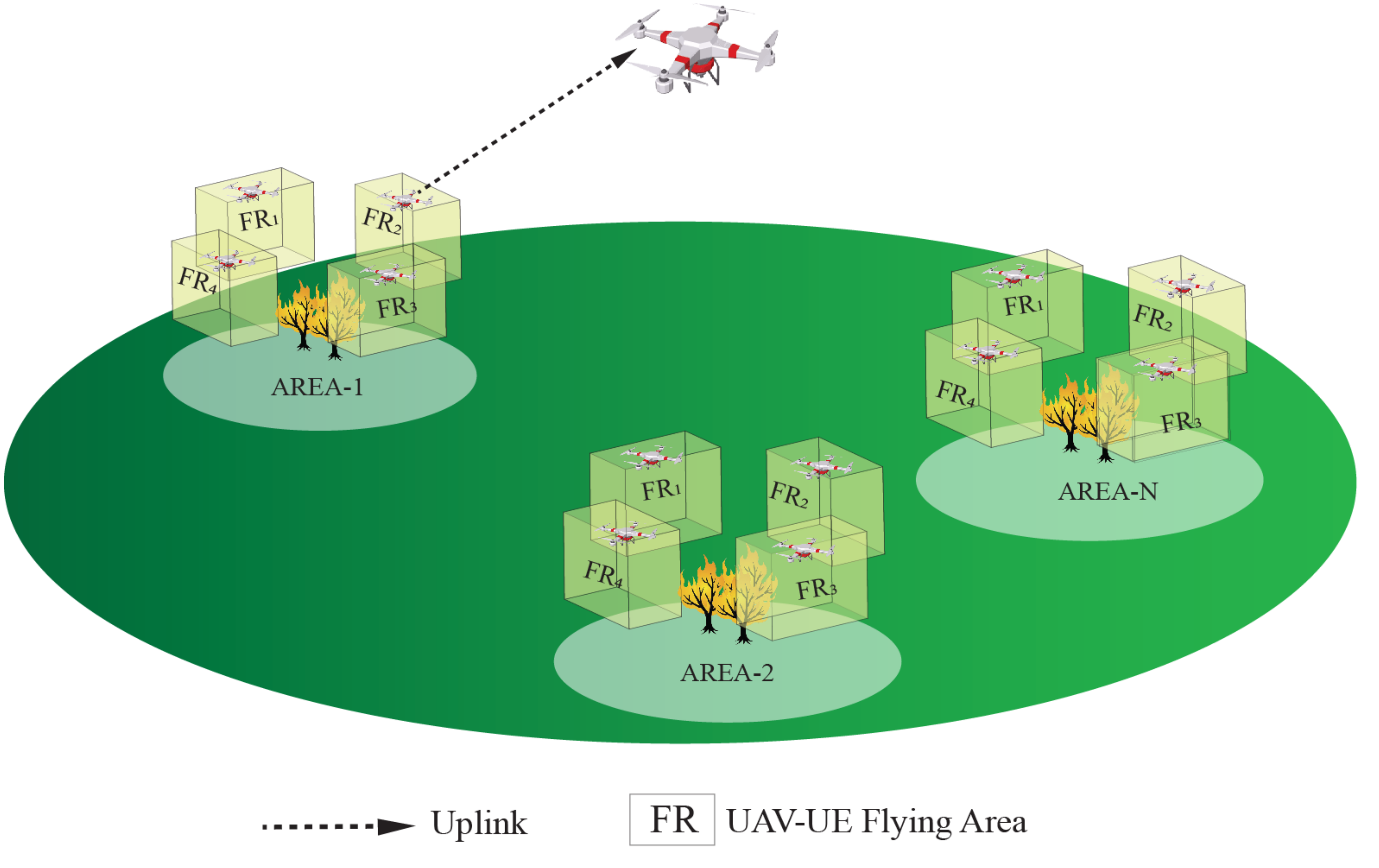}
    \caption{Illustration of System Model}
    \label{fig:model}
\end{figure}

\section{System Model and Problem  Formulation}\label{sectio}
As illustrated in Fig. \ref{fig:model}, we consider a single UAV-BS to provide a network coverage for multiple UAV-UEs to satisfy the network rate requirement of each UAV-UE to stream high quality video of multiple firefighting areas. 
The UAV-BS is located at the center of the environment, such as forest area, with the maximum coverage radius $r_{\text{max}}$. The UAV-BS is connected  through wireless network to the fixed or mobile control station. We assume that the arriving distribution of the fire video streaming request is the same as that of the fire arrival distribution \cite{podur2010fire}, which follows Poisson process distribution with density $\lambda_a$. The UAV-BS receives a request when a fire event  occurs, and the $k$th UAV-UE automatically flies to the center of  $k$th flying region $\text{FR}_k$ to serve the $i$th fire area $A_i (x_{i},y_{i})$.

We consider a video streaming task that lasts for $T$ time slots with an equal duration $t$. The selection of the optimal location to stream the video plays an important role in ensuring the UAV-UEs capture the full firefighting area of $A_i$. Therefore, the $k$th UAV-UE needs to find the optimal position $U (x_{k}^*, y_{k}^*, h_{k}^*)$ to transmit the video to the UAV-BS. The size of the $k$th fire region  $\text{FR}_k$ for the $k$th UAV-UE  depends on the number of UAV-UEs that perform the video streaming for the $i$th fire area $A_i$. To make sure that all UAV-UEs can jointly  capture the panoramic video of $A_i$, $K$ UAV-UEs are distributed evenly around $A_i$, as shown in Fig. \ref{fig:model}. Meanwhile, the UAV-BS also searches for the optimal location $P (x_{BS}^*, y_{BS}^*, h_{BS}^*)$ to satisfy the minimum data rate requirement for all UAV-UEs.
In addition, the safety region of the $A_i$ is considered to guarantee $\text{FR}_k$ and $A_i$, and $A_i$ and $A_{i+1}$ are not overlapping to guarantee that  the UAV-BS and UAV-UEs are safe from fire.

\subsection{Request Arrival}
The request contains the $i$th area  $A_i$ with its centre at $(x_{i},y_{i})$ with radius $r_i$. We assume that $K$ UAV-UEs serve each fire area and stream real-time videos simultaneously. We assume that the height of the fire $h_{i}$  follows Log-normal distribution \cite{val2018global_fire}, thus, the minimum flying height of all UAVs is $h_{\text{min}}$, which satisfy $h_{\text{min}}= \max(h_i)$. All UAV-UEs in $A_i$ will be operated at the same altitude. The environment is divided into $W$ square grids, thus, the length, width and height of each grid are $\frac{X}{\sqrt[3]{W}},\frac{Y}{\sqrt[3]{W}},\frac{Z}{\sqrt[3]{W}}$, respectively. 
At the $t$th time slot, the flying position $ \vec {U}(x_{i,k} , y_{i,k}, h_{i,k})$ of the $k$th UAV-UE can be calculated as
\begin{align} \label{u}
    \vec {U}_{t+1} (x_{i,k} , y_{i,k}, h_{i,k}) =  \vec {U}_{t} (x_{i,k}, y_{i,k}, h_{i,k}) +  \vec {a}_t (x,y,z), 
\end{align}
\newline
with
\newpage
\begin{strip}

\begin{align} \label{FR_k}
    x_i - a  \leq &  x _ {i,k } \leq  x _ { i  } + a, \\
   y_i - a  \leq & y _ {i, k } \leq y _ { i }+ b,  \\
   h_{\text{min}} \leq & h _ { i, k} \leq  h_{\text{max}},
\end{align}

\begin{subequations} \label{FR}
\begin{align} 
    U_{(i,k=1)}^t = \lbrace (x_{1},y_{1},h_{1}) \vert  x _ { i }  - a  \leq   x _ { i , 1 } \leq  x _ { i } +a ,    y _ { i } + a \leq  y _ { i , 1 } \leq y _ { i } + b  , h _ { i }   \leq  h _ { 1} \leq  h _ { max}  \rbrace,  \\
    U_{(i,k=2)}^t = \lbrace (x_{2},y_{2},h_{2}) \vert x _ { i }  - b  \leq   x _ { i , 2 } \leq  x _ { i } -a ,    y _ { i } - a \leq  y _ { i , 2 } \leq y _ { i } + a  , h _ { i }   \leq  h _ { 2} \leq  h _ { max}  \rbrace,  \\
    U_{(i,k=3)}^t = \lbrace (x_{3},y_{3},h_{3}) \vert x _ { i }  - a  \leq   x _ { i , 3 } \leq  x _ { i } +a ,    y _ { i } -b \leq  y _ { i , 3 } \leq y _ { i } -a  , h _ { i }   \leq  h _ { 3} \leq  h _ { max}  \rbrace,  \\
    U_{(i,k=4)}^t = \lbrace (x_{4},y_{4},h_{4}) \vert x_ { i }  + a  \leq   x _ { i , 4 } \leq  x _ { i } +b ,    y _ { i } -a \leq  y _ { i , 4 } \leq y _ { i } + a  , h _ { i }   \leq  h _ { 4} \leq  h _ { max}  \rbrace.  \label{FR4}
\end{align}
\end{subequations}
\end{strip}

\begin{figure}[!t]
    \centering
    \includegraphics[width=3.5 in,height=2.5in]{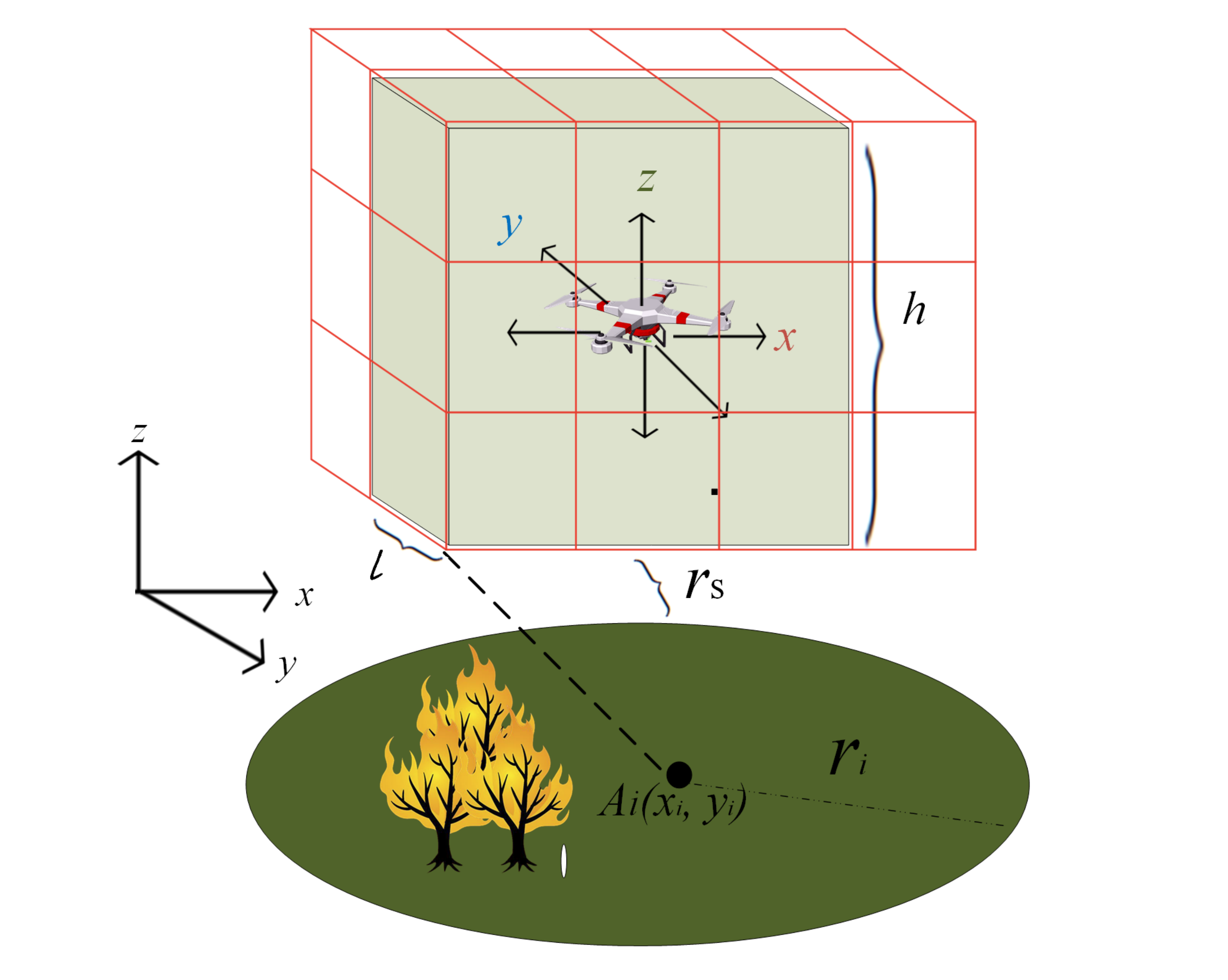}
    \caption{Flying boundry of the $k$th UAV-UE.}
    \label{fig:UE}
\end{figure}
\noindent where $\vec {a}_t (x,y,z)$ is the action vector, $a= r_i +r_s$, $b=r_i +r_s+l$, $r_s$ is the safe distance between $A_i$ and $\text{FR}_k$,  $l$ is the length of flying region, and $h_{\text{max}}$ is the maximum height of UAV-UE regulated by the government (i.e. 120m in UK \cite{gov.uk_2017}). Furthermore, to capture full panoramic video, we propose the boundary flying area for UAV-UEs in each fire area, which can be written as Eq. (\ref{FR}).

\subsection{Channel Model}
\begin{figure} [!t]
    \centering
    \includegraphics[width=3.5 in,height=3.5in]{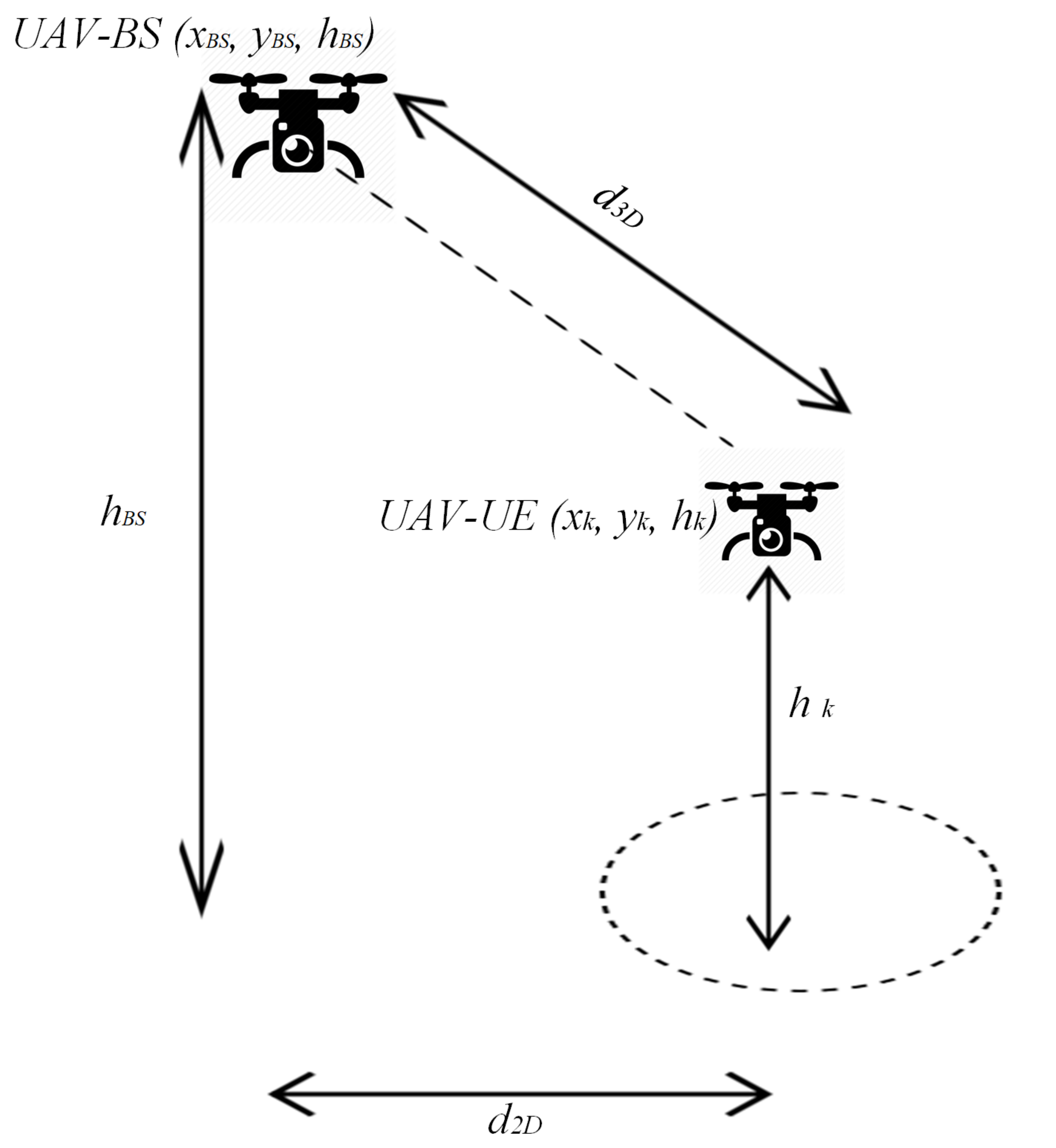}
    \caption{UAV-to-UAV communication.}
    \label{fig:u2u}
\end{figure}

In the wireless network, we assume that the channel model between the $k$th UAV-UE and the UAV-BS  contains large-scale fading (path loss and channel gain) and small-scale fading \cite{azari2019u2u}.
We assume that the link between the UAVs are line-of-sight (LoS). The pathloss from the $k$th UAV-UE  to the UAV-BS can be written as  

\begin{align} \label{PL}
PL_{\mathrm{LoS,k}}(t) &=20 \log \left(\frac{4 \pi f_{c} d^{k}_{3D}(t)}{c}\right)+\eta_{\mathrm{LoS}},  
\end{align}
where $f_c$ is the carrier frequency, $c$ is the speed of light in vacuum, $\eta_{Los}$ is the additional attenuation factors due to the LoS connection, and $d^{k}_{3D}(t)$ is distance between the $k$th UAV-UE and the UAV-BS, as shown in Fig. \ref{fig:u2u}, which can be calculated as
\begin{equation}
\begin{aligned} 
    d_{3D}^{k}(t)=\sqrt{\left(x_{BS}(t)-x_{{k}}(t)\right)^{2}+\left(y_{BS}(t)-y_{{k}}(t)\right)^{2} 
    +\left(h_{BS}(t)-h_{k}(t))\right)^{2}}.
\end{aligned}
\end{equation}

In our model, we use the Rician distribution  \cite{durgin2003space}\cite{goddemeier2015investigation} to  define small scale fading $p_{\xi}({d_k})$, which can be denoted as

\begin{equation}
p_{\xi}(d_k)=\frac{d_k}{\sigma_{0}^{2}} \exp \left(\frac{-{d_k}^{2}-\rho^{2}}{2 \sigma_{0}^{2}}\right) I_{0}\left(\frac{{d_k} \rho}{\sigma_{0}^{2}}\right),
\end{equation}
\noindent with ${d_k}\geq$ 0, and $\rho$ and $\sigma$ are the strength of the dominant
and scattered (non-dominant) paths, respectively. The Rice factor $\kappa$ can be defined as
\begin{equation}
\kappa=\frac{\rho^{2}}{2 \sigma_{0}^{2}}.
\end{equation}

It is possible that the selected position of each UAV-UE can generate more interference to the UAVs nearby, which can result in poor transmission performance and make it difficult for the UAV-UE to maintain the connection with the UAV-BS. Power control can be a solution to minimize the uplink interference among UAV-UEs at appropriate power level \cite{yajnanarayana2018interference}. Through properly controlling the transmit power of each UAV-UE in the uplink transmission, the interference among UAV-UEs can be mitigated.
According to the 3GPP guidelines \cite{3gpp.36.777}, we consider fractional power control for all UAVs and the power transmitted by the $k$th UAV-UE while communicating with the UAV-BS can be given by
\begin{equation}
P_{U_{k}}=\min \left\{P_{U_{k}}^{\max }, \left(\begin{array}{c}
10 \log _{10}\left(B)\right) +\rho_{\mathrm{u_{k}}} P L 
\end{array}\right)\right\},
\end{equation}
where $P_{U_{k}}^{\max}$ is the maximum transmit power of the UAV-UE, $B$ is the channel bandwidth, and $ \rho_{\mathrm{u_{k}}}$ = $\lbrace {0, 0.4, 0.5, 0.6, 0.7, 0.8, 0.9, 1} \rbrace$  is a
fractional path loss compensation power control parameter \cite{yajnanarayana2018interference}.

In the proposed wireless UAV network, the received power from the $k$th UAV-UE to the UAV-BS at the $t$th time slot is presented as

\begin{align} \label{power}
P_{k}(t)&=P_{U_k} G\left(d_{3D}(t)\right)^{-\alpha} 10 ^{\frac{-p_{\xi}({d_k})}{10}},
\end{align}
where $P_{U_k}$ is the transmit power of the $k$th UAV-UE, $G$ is channel power gains factor introduced by amplifier and antenna \cite{zhang2019u2x}, $\left(d_{3D}(t)\right)^{-\alpha}$ is the pathloss, $\alpha$ is the path loss exponent, and $p_{\xi}({d_k})$ is the Rician small scale fading. 
The interference from the $m$th UAV-UE to the UAV-BS at the $t$th time slot can be written as
\begin{equation}
I_{ U 2 U}(t)=\sum_{\mathbf{m} \in \mathbf{K} \backslash {k}} \psi_{ m}(t) P_{m }(t),
\end{equation}
where $\psi_m(t)=1$ indicates that the transmission between the $k$th UAV-UE and the UAV-BS is active, otherwise, $\psi_m(t)=0$, and $ P_{m }(t)$ is the transmit power of $m$th UAV-UE.
The signal to interference plus noise ratio (SINR) of the UAV-BS is given by
\begin{align}
    \gamma_{k}(t)=\frac{P_{k}(t)}{N + \sum_{\mathbf{m} \in \mathbf{K} \backslash {k}} \psi_{ m}(t) P_{m }(t)},
\end{align}
where $N$ is the noise power at the UAV-BS whose elements are average of independent random Gaussian variables with the variances
$\sigma_n^{2}$. Then, the transmission uplink rate from the $k$th UAV-UE to the UAV-BS can be denoted as
\begin{equation}\label{rate}
R_{k}(t)= B \log _{2}\left(1+\gamma_{k}(t) \right).
\end{equation}

\begin{table*}[!t]
\centering
\caption{Type of Video Quality \cite{quality_video}}
\label{tab:quality}
\begin{tabular}{|c|c|c|c|c|c|}
\hline
Video Quality & Resolution (pixels) & Framrate (FPS) & Bitrate (average) & Data used per minute & Data used per 60 minutes \\ \hline
144p          & 256x144             & 30             & 80-100 Kbps       & 0.5-1.5 MB           & 30-90 MB                 \\ \hline
240p          & 426x240             & 30             & 300-700 Kbps      & 3-4.5 MB             & 180-250 MB               \\ \hline
360p          & 640x360             & 30             & 400-1,000 Kbps    & 5-7.5 MB             & 300-450 MB               \\ \hline
480p          & 854x480             & 30             & 500-2,000 Kbps    & 8-11 MB              & 480-660 MB               \\ \hline
720p (HD)     & 1280x720            & 30-60          & 1.5-6.0 Mbps      & 20-45 MB             & 1.2-2.7 GB               \\ \hline
1080p (FHD)   & 1920x1080           & 30-60          & 3.0-9.0 Mbps      & 50-68 MB             & 2.5-4.1 GB               \\ \hline
\end{tabular}
\end{table*}

\subsection{Video Streaming Model}
In this paper, we consider the long-term video streaming that are modelled as consecutive video segments. Each segment consists  of multiple frames, and the frame is considered to be the smallest data unit. The resolution of each frame corresponds to its minimum data rate requirement. 
Table \ref{tab:quality} presents the type of Video Quality \cite{quality_video}. 
For example, if the communication rate (bitrate) is between 300-700 kbps, the video type that we should consider to use is 240 p. Knowing that  144p corresponds to the  smallest size of the video type, all UAV-UEs need to satisfy the minimum uplink bitrate, i.e., $R_{min}$=80 kbps. 

Each UAV-UE is equipped with a nonordinary optical camera with the resolution of $r_{px} \times r_{py}$, and the video is consisted of multiple consecutive frames \cite{govil2020preliminary_fire}, which is used to monitor the fire area with three main goals: 1) detect the size of fire by continuous capturing  the panoramic video; 2) verify and locate fires reported; and 3) closely monitor a known fire by streams using distribution relationship around the incident.  The quality of the video frame depends on its resolution of the $i$th video frame at the $t$th time slot $v_i (t)$. Furthermore, for each video frame, we assume that it has the same playback time $T_l$, i.e. 2ms to 4ms, which depends on 30 FPS or 60 FPS. In addition, the delay of video streaming via UAVs is consisted of three elements, i.e. capture time, encoding time, and transmission time. As all UAVs capture a video using the same resolution, the capture time and the encoding time are constant. Thus, we mainly focus on the uplink transmission time, which can be expressed as
\begin{equation}
T_{i, k}(t)=\frac{D(v_{i}(t))}{R_k(t)}=\frac{r_{p x} \cdot r_{p y} \cdot b}{B \log _{2}\left(1+\gamma_{k}(t) \right)},
\end{equation}
where $b$ is the number of bits per pixel, and $D(v_i(t))$ is the data size based on $v_i(t)$. The video frames are processed in parallel in multi-core processors, and the time consumption at the $t$th time slot is $T(t)= \max \lbrace{T_{i,k}(t)}\rbrace$ \cite{carballeira2012framework_parallel}. To guarantee the smoothness and seamless of the video streaming, $T(t)$ must satisfy the delay constraint, namely, $T(t) < T_l$.

\subsection{Quality of Experience Model}
The key parameters of video streaming  are video quality, quality of variation, rebuffer time, and the startup delay \cite{yin2015streaming}. According to \cite{xiao2019bitrate}, the rebuffering time and startup delay can be ignored. Thus,
the video transmission may be suffered from a delay, which can be calculated as $D(t) = T(t)-T_l$, with $T_l$ as the delay constraint. The QoE can be formulated as the sum of QoE over all the areas and all the UAV users, and denoted as 
\begin{equation}
\begin{aligned} \label{QoE}
    QoE =&\frac{\kappa_{i,k}(t)}{IK} \biggl  (\sum_{i=1}^{I} \sum_{k=1}^{K}   q(R_{i,k}(t)) \\ &-  |q(R_{i,k}(t))-q(R_{i,k}(t-1))|  \biggr) - \omega(t) D(t),
\end{aligned}
\end{equation}
where $q(R_{i,k}(t))$ is video quality metrics \cite{Mao2017}, which can be written as
\begin{align}
    q(R_{i,k}(t))= \log \left( \frac{R_{i,k}(t)}{R_{\text{min}}(v_i(t))}\right),
\end{align}
$\kappa_{i,k}(t)$ and $\omega(t)$ are the weight of video quality and delay, respectively. As our aim is to maximize the QoE, the condition of $\kappa_{i,k}(t) > \omega (t)$ must be guaranteed, and $R_{\text{min}}(v_i(t))$ is the minimum rate that should be satisfied for the selected $v_i(t)$.

\subsection{Problem Formulation}
Our aim is to maximize the QoE that jointly exploit the optimal positions of the UAV-BS and UAV-UEs, and the optimal adaptive bitrate selection. The fluctuation of the transmission link will cause unstable network performance that leads to low QoE and high delay. Thus, to minimize the delay at each Transmission Time Interval (TTI) and maximize the quality of video streaming, we jointly consider the optimal UAV-BS location $\mathcal{P}=(x_{BS}(t), y_{BS}(t), h_{BS}(t))$, the position of the $k$th UAV-UE $\mathcal{U} = (x_{i,k}(t), y_{i,k}(t), h_{i,k}(t))$, the maximum power of UAV-UE $P_{U_{k}}$, and the bitrate resolution $\mathcal{V} = \lbrace 144, 240, 360, 480, 720, \text{and}~1080 \rbrace  $ p. The optimization problem can be formulated as
\begin{equation}
\begin{aligned}  \label{opti}
    \max _{\left\{\mathcal{P},  \mathcal{U}, P_{U_{k}},  \mathcal{V}\right\}} 
    &\frac{\kappa_{i,k}(t)}{IK} \biggl  (\sum_{i=1}^{I}\sum_{k=1}^{K}   q(R_{i,k}(t) \\ &-  |q(R_{i,k}(t))-q(R_{i,k}(t-1))|)  \biggr) - \omega(t) D(t),
\end{aligned}
\end{equation}
\text {s.t.} 
\begin{align}
&\max h_i > h_{BS} (t) >  h_{\max},  \label{height} \\
&R_{i,k} (t)  > R_{(\text{min})}^k(v_i(t)), \label{min_rate}\\
&\sqrt{(x_{BS}(t)-x_i )^2+(y_{BS} (t)-y_i )^2} > r_i + r_s,  \label{safety}\\
&\mathcal{U} \in \text{Eq}.  \eqref{u}. 
\end{align}

The objective function in Eq. (\ref{opti}) captures the average QoE received at the UAV-BS. 
The UAV-BS's height must follow the condition in  Eq.  (\ref{height}). Eq. (\ref{min_rate}) guarantees $R_k$ obtained from $\mathcal{U}_k$ to meet the minimum requirement of data rate of UAV-UEs based on the adaptive bitrate selection.
Then, Eq. (\ref{safety}) guarantees that the position of the UAV-BS will not intersect with the UAV-UE's flying region. $\mathcal{U}$ follows the requirement of the flying region $\text{FR}_i$ presented in Eq. (1). In the experiment, the UAVs are hover and flying at constant speed.


\section{ Optimization Problem via Reinforcement Learning} \label{method}
In this section, we design several DRL algorithms to solve QoE maximization problem in UAV-to-UAV network and to be compared with existing traditional method - Greedy algorithm. Specifically,  we propose two DRL algorithms, which are Deep Q-Learning and Actor-Critic, to maximize the QoE of live video streaming in U2U communication. 
\subsection{Reinforcement Learning}
For our proposed RL-based method, the UAV-BS acts as an agent to collect video from UAV-UEs while maximizing Quality of Experience (QoE). The QoE optimization problem is influenced  by the delay, UAVs' positions, and bitrate selection during each Transmission Time Interval (TTI), and forms a Partially Observed Markov Decision Problem (POMDP). Through learning algorithms, the UAV-BS is able to select the positions of the UAV-BS $\mathcal{P}$, the UAV-UEs $\mathcal{U}$, and the adaptive resolution $\mathcal{V}$, in order to maximize the QoE.

\subsubsection{State Representation}  The current state $s(t)$ corresponds to a set of current observed information. The state of the UAV-BS can be denoted as $s=[ \mathcal{P}, \mathcal{V}, \mathcal{U}, \text{QoE}]$, where $\mathcal{P}$= ($x_{BS}(t)$, $y_{BS}(t)$, $h_{BS}(t)$) is the position of the UAV-BS, $\mathcal{V}$ is the bitrate selection, and $\mathcal{U} = (x_{k}(t), y_{k}(t), h_{k}(t))$ is the  positions of UAV-UEs.

\subsubsection{Action Space} 
Q-agent will choose action $a =({BP,BU,BV})$ from set $\mathcal{A} $. The dimension of the action set can be calculated as $\mathcal{A} = BP \times BU^{i \times k} \times BV^i \times P $.  The actions for UAVs include (i) UAV-BS's flying direction (BP), (ii) UAV-UEs' flying direction (BU), (iii)  resolution of the $i$th UAV-UE (BV),  and (iv) UAV-UE's power (P). The action space is presented as
\begin{itemize}
    \item   BP = (Position coordinate follows Eq.(21) )
    \item BU  = (Position coordinate with boundaries of Eq.(22))
    \item BV= (144, 240, 360, 480, 720, or 1080) p
    \item P = (23, 25, 30) dBm
\end{itemize}

To ensure the balance of exploration and exploitation actions of the UAV-BS, $\epsilon$ -greedy ( $0<\epsilon \leq 1$) exploration is deployed. At the $t$th TTI, the UAV-BS randomly generates a probability $p_{\epsilon}({t})$ to compare with $\epsilon $. If the probability $p_{\epsilon}({t}) < \epsilon $, the algorithm randomly selects an action from the feasible actions to improve the value of the non-greedy action. However, if $ p_{\epsilon}({t}) \geq \epsilon $, the algorithm exploits the current knowledge of the Q-value table to choose the action that maximizes the expected reward.

\subsubsection{Rewards}
When the $a({t})$  is performed, the corresponding reward $\text{re}({t})$ is defined as
\begin{equation}
\begin{aligned} \label{QoE}
    {re}(t) =&\frac{\psi_{i,k}(t)}{IK} \biggl  (\sum_{i=1}^{I}\sum_{k=1}^{K}   q(R_{i,k}(t)) \\ &-  |q(R_{i,k}(t))-q(R_{i,k}(t-1))|  \biggr) - \omega(t) D(t),
\end{aligned}
\end{equation}
where $q(R_{i,k}(t))$ is video quality metrics \cite{Mao2017}, which can be written as
\begin{align}
    q(R_{i,k}(t))= \log \left( \frac{R_{i,k}(t)}{R_{\text{min}}(v_i(t))}\right),
\end{align}
$\psi_{i,k}(t)$ and $\omega(t)$ are the weights of video quality and delay, respectively. If $R_{i,k}(t)$ is unable to satisfy the minimum transmission rate for $R_{\text{min}}^k(v_i(t))$, namely, $R_{i,k} (t)  < R_{\text{min}}^k(v_i(t))$, the system will receive negative reward, which means $\text{re}(t) <0 $.

\subsection{Q-learning}
The learning algorithm needs to use Q-table to store the state-action values according to different states and actions. Through the policy $\pi (s, a) $, a value function $Q(s,a)$ can be obtained through performing action based on the current state. At the $t$th time slot, according to the observed state $s(t)$, an action $a(t)$ is selected following  $\epsilon$ -greedy approach from all actions. By obtaining a reward $\text{re}(t)$, the agent updates its policy $\pi$ of action $a(t)$. Meanwhile, Bellman Equation is used to update the state-action value function, which can be denoted as
\begin{equation}\label{bellman}
\begin{aligned} 
Q(s(t), a(t))=&(1-\alpha) Q(s(t), a(t))\\& +\alpha\left\{ \text{re}({t+1})+\gamma \max _{a(t) \in \mathcal{A}} Q(s({t+1}), a(t))\right\}, \end{aligned}
\end{equation}
where $\alpha$ is the learning rate, $\gamma\in [0,1)$ is the discount rate that determines how current reward affects the updating value function. Particularly, $\alpha$ is suggested to be set to a small value (e.g., $\alpha$ = 0.01) to guarantee the stable convergence of training. 

\subsection{Deep Q-learning}
However, the dimension of both state space and action space can be very large if we use the
traditional tabular Q-learning, which will cause high computation complexity. To solve this problem, deep learning is combined with Q-learning, namely, 
Deep Q-Network (DQN), where a deep neural network (DNN) is used to approximate the state-action value function. $Q(s, a)$ is parameterized by using a function $Q(s, a; \boldsymbol{\theta}_{\text{DQN}})$, where  $\boldsymbol{\theta}_{\text{DQN}}$  is the weight matrix of DNN with multiple layers. $s$ is the state observed by the UAV and  acts as an input to Neural Networks (NNs).
The output are selected actions in ${\mathcal{A}}$. Furthermore, the intermediate layer contains multiple hidden layers and is connected with Rectifier Linear Units (ReLu) via using $f(x)= \max (0,x)$ function. At the $t$th time slot, the weight vector is updated by using Stochastic Gradient Descent (SGD) and Adam Optimizer, which can be written as

\begin{equation}
\boldsymbol{\theta}_{\text{DQN}}({t+1})=\boldsymbol{\theta}_{\text{DQN}}(t)-\lambda_{\text{ADAM}} \cdot \nabla \mathcal{L}(\boldsymbol{\theta}_{\text{DQN}}(t)),
\end{equation}
where $\lambda_{\text{ADAM}}$ is the Adam learning rate, and $\lambda_{\text{ADAM}} \cdot \nabla \mathcal{L}(\boldsymbol{\theta}_{\text{DQN}}(t))$ is the gradient of the loss function $\mathcal{L}(\boldsymbol{\theta}_{\text{DQN}}(t))$, which can be written as
\begin{equation}\label{GDS}
\begin{array}{c}
\nabla \mathcal{L}\left(\theta_{\mathrm{DQN}}(t)\right)=\mathbb{E}_{S^{i}, A^{i}, \mathrm{re}(i+1), S^{i+1}}\left[\left(Q_{\mathrm{tar}}-Q\left(S^{i}, A^{i} ;\right.\right.\right. \\
\left.\left.\theta_{\mathrm{DQN}}(t)\right) \cdot \nabla Q\left(S^{i}, A^{i} ; \theta_{\mathrm{DQN}}(t)\right)\right]
\end{array}
\end{equation}

where the expectation is calculated with respect to a so-called minibatch, which are randomly selected in previous samples $(S^{i}, A^{i},  {Re}^{i+1}, S^{i+1})$ for some $i \in\left\{t-M_{r}, t-M_{r}+1, \ldots, t\right\}$, with $M_{r}$ being the replay memory. The minibatch sampling is able to improve the convergence reliability of the updated value function \cite{mnih2015human_rl}. In addition, the target Q-value $Q_{\text{tar}}$ can be estimated by
\begin{equation}
Q_{\text{tar}}= {re}^{i+1}+\gamma \max _{a \in \mathcal{A}} Q(S^{i+1}, a ; \boldsymbol{\bar{\theta}}_{\text{DQN}}({t})),
\end{equation}
where $\boldsymbol{\bar{\theta}}_{\text{DQN}}(t)$ is the weight vector of the target Q-network to be used
to estimate the future value of the Q-function in the update rule. This parameter is periodically copied from the current value $\boldsymbol{\theta}_{\text{DQN}}(t)$ and kept fixed for a number of episodes. The DQN algorithm is presented in Algorithm 1.

\begin{algorithm}[!t]
\caption{: Optimization by using DQN}

Input: {The set of UAV-BS position $\lbrace x_{BS}, y_{BS}, h_{BS} \rbrace$}, bitrate selection $V$, the position of the $k$th UAV-UE  $U_k = (x_{k}^t, y_{k}^t, h_{k}^t)$, $\sum QoE$ and operation iteration $I$. \\
 \textbf {Algorithm hyperparameters:} Learning rate $\alpha \in (0,1]$, $\epsilon \in (0,1]$, target network update frequency $K$;\\
Initialization of replay memory $M$, the primary Q-network $\boldsymbol{\theta}$, and the target Q-network $\boldsymbol{\bar\theta}$;\\
    \textbf{For} {$e \leftarrow 1$ \textbf{to} $I$}{\\
    \hspace{0.7cm} Initialization of $s^1$ by executing a random action $a^0$;\\
\textbf{For}{$t \leftarrow 1$ \textbf{to} $T$ }{\\
	\hspace{0.7cm} \textbf{If} {$p_{\epsilon} < \epsilon$} } 
	Randomly select action $a^t$ from ${\mathcal{A}}$;\\
    \hspace{0.7cm} \textbf{else}  select $a^{t}=\underset{a \in \mathcal{A}}{\operatorname{argmax}} Q\left(S^{t}, a, \theta\right)$;\\
	\hspace{0.5cm} The UAV-BS performs $a^t$ at the $t$th TTI ;\\
	\hspace{0.5cm} The UAV-BS observes $s^{t+1}$, and calculate $re^{t+1}$ using Eq. \eqref{QoE};\\
	\hspace{0.5cm} Store transition $(s^t; a^t; re^{t+1}; s^{t+1})$ in replay memory $M$;\\
	\hspace{0.3cm} Sample random minibatch of transitions $(S^i ;A^i ;Re^{i+1}; S^{i+1})$ from replay memory $M$;\\
	\hspace{0.5cm} Perform a gradient descent for $Q(s; a; \boldsymbol{\theta})$ using \eqref{GDS} ;\\
	\hspace{0.5cm} Every $K$ steps update target Q-network $\boldsymbol{\bar\theta}$ = $\boldsymbol{\theta}$}.\\

\end{algorithm}

\subsection{Actor-Critic}
Different from the DQN algorithm, which obtains the optimal strategy indirectly by optimizing the state-action value function, the AC algorithm directly determines the strategy that should be executed by observing the environment state.
The AC algorithm combines the advantages of value-based function method and policy-based function method. In the AC algorithm, the agent is consisted of two parts, i.e., actor network and critic network, and it solves the problem through using two neural networks. Meanwhile, the AC algorithm deploys a separate memory structure to explicitly represent the policy which is independent of the value function. The policy structure is known as the actor network, which is used to select actions. Meanwhile, the estimated value function is known as the critic network, which is used to criticize the actions performed by the actor. The AC algorithm is an on-policy method and temporal difference (TD) error is deployed in the critic network. To sum up, the actor network aims to improve the current policies while the critic network evaluates the current policy to improve the actor network in learning process. 

The critic network uses value-based learning to learn a value function. The state-action value function $V(s(t), \boldsymbol{w}(t))$ in the critic network can be denoted as
\begin{equation}
V(s, \boldsymbol{w}(t))=\boldsymbol{w}^{\top}(t) \boldsymbol{\Phi}(s(t)),
\end{equation}
where $\boldsymbol{\Phi}(s(t)) = s(t)$ is state features vector and $\boldsymbol{w}(t)$ is critic parameters, which can be updated as
\begin{equation}
\boldsymbol{w}(t+1)=\boldsymbol{w}(t)+\alpha_c(t) \delta(t) \nabla_{\boldsymbol{w}} V\left(s(t), \boldsymbol{w}(t)\right),
\end{equation}
where $\alpha_c$ is the learning rate in the critic network. After performing the selected action,  TD error $\delta(t)$ is used to evaluate whether the selected action based on the current state performs well \cite{zhang2019qoe}, which can be calculated as
\begin{equation}
\delta(t)= \text{re}({t+1})+\gamma_{\boldsymbol{w}} (V\left(s({t+1}), \boldsymbol{w}(t)\right)-V\left(s({t}), \boldsymbol{w}(t)\right)).
\end{equation}
Then, the actor network is used to search the best policy to maximize the expected reward under the given policy with parameters $\boldsymbol{\theta}_{\text{AC}}$, which can be updated as
\begin{equation}
\boldsymbol{\theta}_{\text{AC}}(t+1)=\boldsymbol{\theta}_{\text{AC}}(t)+\alpha_a \nabla_{\boldsymbol{\theta}_{\text{AC}}} J\left(\pi_{\boldsymbol{\theta}_{\text{AC}}(t)}\right),
\end{equation}
where $\alpha_a$ is the learning rate in the actor network, which is positive and must be small enough to avoid causing oscillatory behavior in the policy, and according to \cite{zhang2019qoe},  $\nabla_{\boldsymbol{\theta}_{AC}} J\left(\pi_{\boldsymbol{\theta}_{AC}}\right)$ can be calculated as  
\begin{equation}
    \nabla_{\boldsymbol{\theta}_{\text{AC}}} J\left(\pi_{\boldsymbol{\theta}_{\text{AC}}(t)}\right) = \delta(t) \nabla_{\boldsymbol{\theta}_{\text{AC}}} \ln \left(\pi\left(a_t| s_{t}, \boldsymbol{\theta}_{\text{AC}}(t)\right)\right).
\end{equation}
The AC algorithm is presented in Algorithm 2.
\begin{algorithm}[!t]
\caption{: Actor-Critic Algorithm}
Inputs:  The set of UAV-BS position $\lbrace x_{BS}, y_{BS}, h_{BS} \rbrace$, bitrate selection $V$, the position of the $k$th UAV-UE $U_k = (x_{k}^t, y_{k}^t, h_{k}^t)$,  $\sum QoE$ and operation iteration $I$. \\
\textbf{Algorithm hyper-parameter}: Learning rate $\alpha_c \in (0,1]$, $\epsilon \in (0,1]$, Target network update frequency $K$;\\
Initialization of policy parameter $\theta_{AC}$,  weight of the actor network $\textbf{w}$,  value of the critic network $\boldsymbol{V}$;\\
\textbf{For} {$e \leftarrow 1$ \textbf{to} $I$}{\\
    \hspace{0.5cm} Initialization of $s^0$ by executing a random action;
	\hspace{5cm} \textbf{For}{$t \leftarrow 1$ \textbf{to} $T$}{\\
	 Select action $a^t$ according to the current policy;\\
	 The UAV-BS observes $s^{t+1}$, and calculate $re^{t+1}$ using \eqref{QoE};\\
 Store transition $(s^t;a^t;re^{t+1}; s^{t+1})$;\\
 Update TD-error functions;\\
 Update the weights $\textbf{w}$ of critic network by minimizing the loss;\\
 Update the policy parameter vector $\theta$ for actor network;\\
 Update the policy $\theta_{AC}$ and state-value function $V(s(t), \boldsymbol{w}(t))$.\\
	}
	}
\end{algorithm} 

\section{Simulation Results} \label{sim}
In this section, we evaluate  our proposed learning algorithms in our problem setup. The area of the region is 5000 m x 5000m x 100m. In the simulation, the maximum flying height $h_{\text{max}}$ of the UAV-BS is 100m, which is satisfied with the maximum flying height 120m that is stipulated by  the UK government. 
We assume that the available video bitrates of the adaptive video streaming for each video frame are  $(80, 300, 700, 1000, 2000, 3000)$ kbps. The target area is captured by $K$ UAV-UE(s),  i.e.,   $K = $ 4 in the $i$th fire area $A_i$ $(i=1, 2, 3, 4, \text{and}, 5)$. 
At the beginning, the UAV-BS will be deployed at the centre of the environment, i.e. (1250, 1250, $h_{\text{min}}$), where $h_{\text{min}}$ is the maximum height of the fire.
When the fire occurs at the remote area, the UAV-UEs will immediately reach the fire location to stream and oversee the real-time situation. 
The height of the UAV-UEs in each fire area are fixed and follow the distribution of the fire height \cite{podur2010fire}. 
The network parameters for the system are shown in Table \ref{tab:parameter} and follow the existing approach and 3GPP  specifications in \cite{zhang2019u2x}, \cite{3gpp.36.777}, and \cite{al2014optimalLAP}. The performance of all results is obtained by averaging around 100 episodes, where each episode is consisted of 100 TTIs. Finally, the channel model parameters and grid environment parameters are set according to \cite{zhang2019u2x}.

\begin{table}[!t]
\centering
\caption{Parameter}
\label{tab:parameter}
\begin{tabular}{lcll}
\cline{1-2}
\multicolumn{1}{|l|}{Parameter}               & \multicolumn{1}{c|}{Value}    &  &  \\ \cline{1-2}
\multicolumn{1}{|l|}{Number of UAV-UEs}       & \multicolumn{1}{c|}{12}       &  &  \\ \cline{1-2}
\multicolumn{1}{|l|}{Transmission power, $PUe$} & \multicolumn{1}{c|}{23 dBm \cite{zhang2019u2x}}   &  &  \\ \cline{1-2}
\multicolumn{1}{|l|}{Bandwidth, $B$} & \multicolumn{1}{c|}{3 MHz }   &  &  \\ \cline{1-2}
\multicolumn{1}{|l|}{Noise variance $\sigma^2$ }       & \multicolumn{1}{c|}{-96 dBm \cite{zhang2019u2x}}   &  &  \\ \cline{1-2}
\multicolumn{1}{|l|}{Center frequency, $f_c$}    & \multicolumn{1}{c|}{2 GHz \cite[pp. 3777]{urllc_uav}}    &  &  \\ \cline{1-2}
\multicolumn{1}{|l|}{Power gains factor, $G$}    & \multicolumn{1}{c|}{-31.5 dB \cite{zhang2019u2x}}  &  &  \\ \cline{1-2}
\multicolumn{1}{|l|}{Alpha, $\alpha$}                   & \multicolumn{1}{c|}{2}        &  &  \\ \cline{1-2}
\multicolumn{1}{|l|}{Channel parameter, $\eta_{LoS}$}  & \multicolumn{1}{c|}{0.1 \cite[pp. 572]{al2014optimalLAP}}        &  &  \\ \cline{1-2}
\multicolumn{1}{|l|}{Channel parameter, $\eta_{NLoS}$} & \multicolumn{1}{c|}{21 \cite{al2014optimalLAP}}       &  &  \\ \cline{1-2}
\multicolumn{1}{|l|}{Channel parameter, $a$} & \multicolumn{1}{c|}{4.88  \cite[pp. 3777]{urllc_uav},\cite[pp. 7]{al2014modelingPL}}       &  &  \\ \cline{1-2}
\multicolumn{1}{|l|}{Channel parameter, $b$} & \multicolumn{1}{c|}{0.43  \cite[pp. 3777]{urllc_uav} ,\cite[pp. 7]{al2014modelingPL}}       &  &  \\ \cline{1-2}
\multicolumn{1}{|l|} {Radius of target region}    & \multicolumn{1}{c|}{1250 m}        &  &  \\ \cline{1-2} \multicolumn{1}{|l|} {Radius of Surveillance region, $r_{i}$}    & \multicolumn{1}{c|}{250 m}        &  &  \\ \cline{1-2}
                                              
                                              & \multicolumn{1}{l}{}          &  & 
\end{tabular}
\end{table}

\begin{table}[!t]
\centering
\caption{Hyperparameter}
\label{tab:hyperparameter}
\begin{tabular}{|l|c|}
\hline
\textbf{Hyperparameter} & \textbf{Value}          \\ \hline
Learning Rate           & 0.1, 0.01    \\ \hline
Initial Exploration     & 1                           \\ \hline
Final Exploration       & 0.1                 \\ \hline
Discount Rate           & 0.8        \\ \hline
Replay memory           & 10000               \\ \hline
\end{tabular}
\end{table}

\begin{figure}[!t]
    \centering
    \includegraphics[width=3.5 in,height=2.5in]{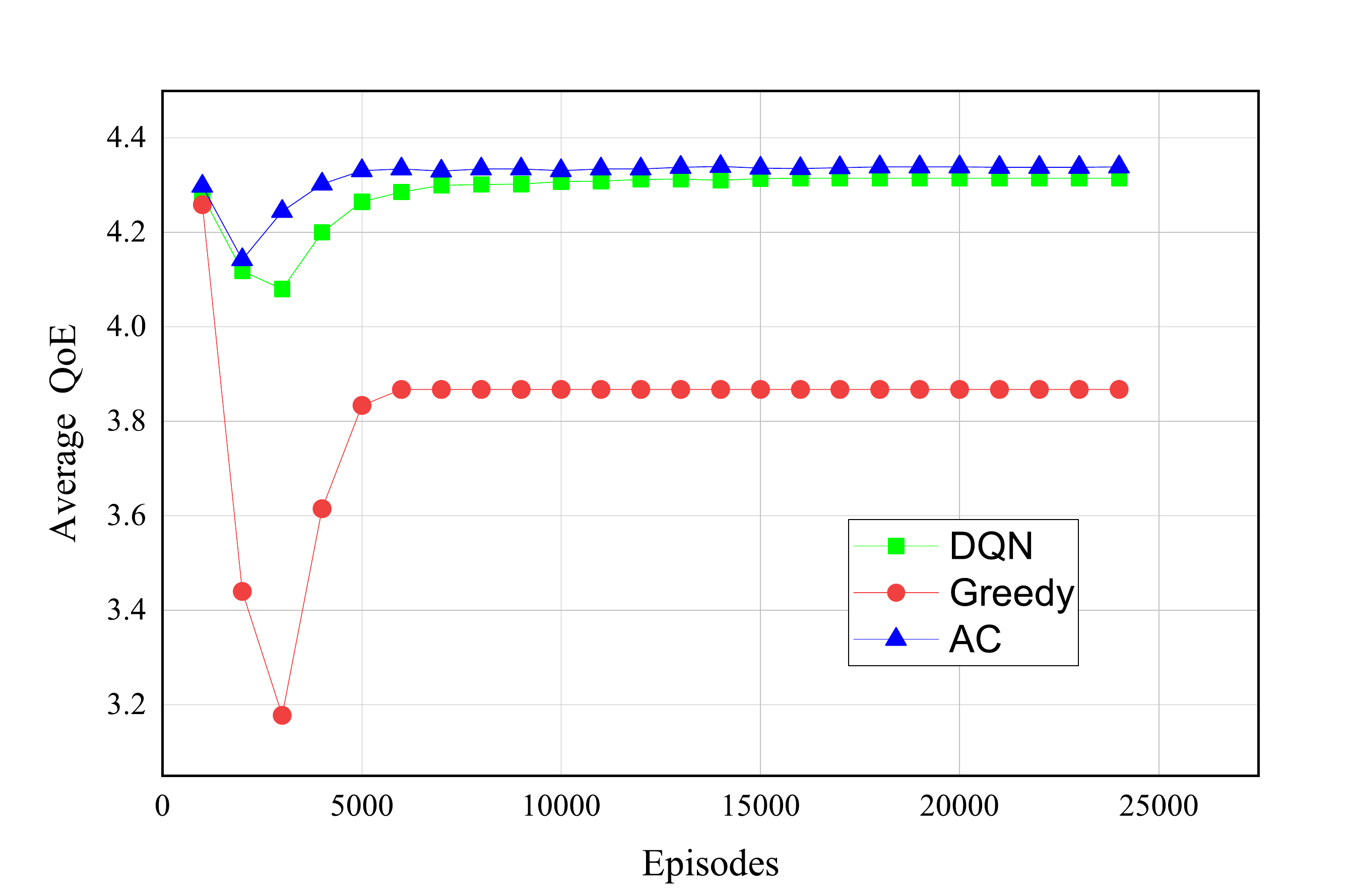}
    \caption{Average QoE value for each frame via
AC, DQN and Greedy algorithms.}
    \label{fig:reward-powercontrol}
\end{figure}

In each scenario, our proposed DQN and AC algorithms are compared with the Greedy algorithm. The Greedy algorithm selects the actions based on the immediate reward and local optimum strategy. The DQN is designed with 3 hidden layers, where each layer consists of 256, 128, 128 ReLU units, respectively. For the AC method, the critic DNN consists  of an input layer with 19 neurons, a fully-connected neural network with two hidden layers, each with 128 neurons, and an output layer with 1 neuron. The UAV-BS is initially set  at the centre of the environment with the height $h_{\text{min}}$. 
In wildfires environment problem, the network coverage with smooth streaming needs to overview the real-time situation. To guarantee high quality of video transmission from multiple UAVs in continuous time slots, the Recurrent Neural Network (RNN) is deployed. 

\begin{figure}[!t]
    \centering
    \includegraphics[width=3.5 in,height=2.5in]{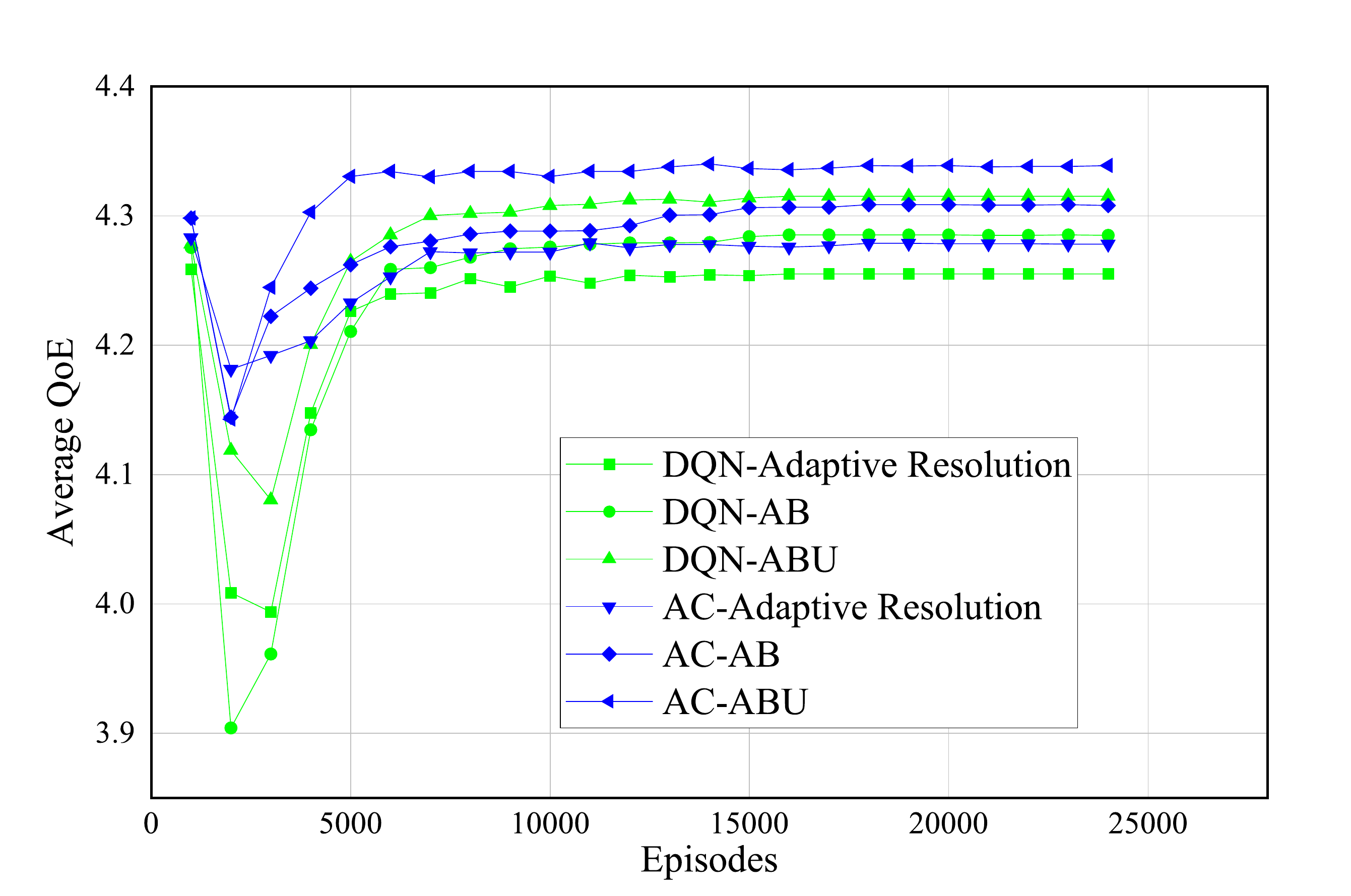}
    \caption{Average QoE of the UAV-BS with different
schemes via different learning algorithms and with different optimization schemes of each episode.}
    \label{fig:reward}
 \end{figure}

\begin{figure}[!t]
    \centering
    \includegraphics[width=3.5in,height=2.5in]{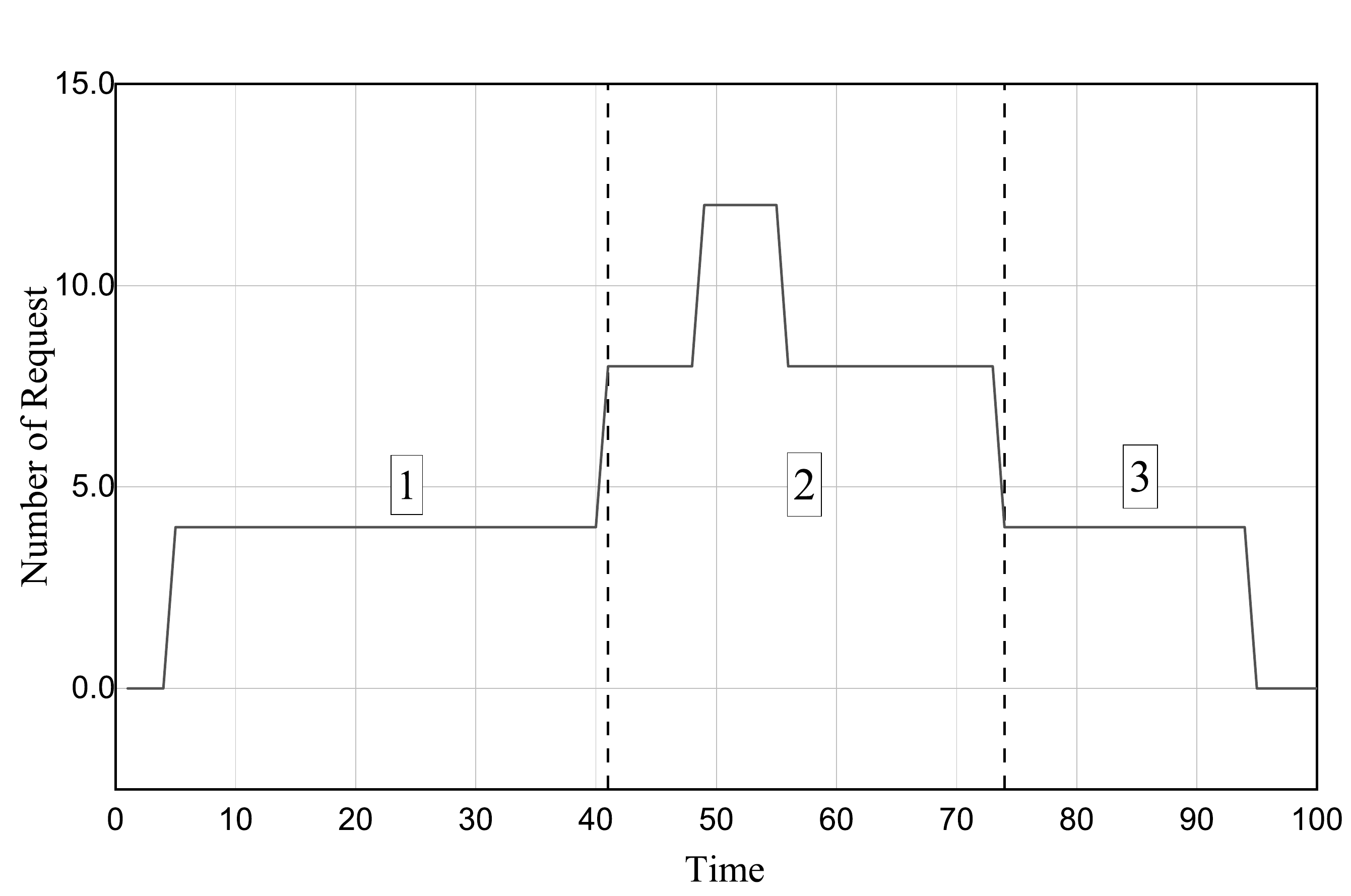}
    \caption{The request of the UAV-UEs in continuous time slots.}
    \label{fig:request}
\end{figure}

Fig. \ref{fig:reward-powercontrol} plots the average  QoE value over all frames via AC, DQN and Greedy algorithms.  It can be seen that DRL algorithms outperform the non-learning based algorithm, i.e., Greedy algorithm. 
Moreover, the convergence speed of the DRL algorithms is faster than that of the Greedy algorithm. Specifically, in the Greedy algorithm, the UAVs only consider exploiting the current reward, rather than exploring the long-term reward. Therefore, the UAVs are not able to achieve higher expected reward  compared to the DRL algorithm.

\begin{figure}[!t]
    \centering
    \includegraphics[width=3.5 in,height=2.5in]{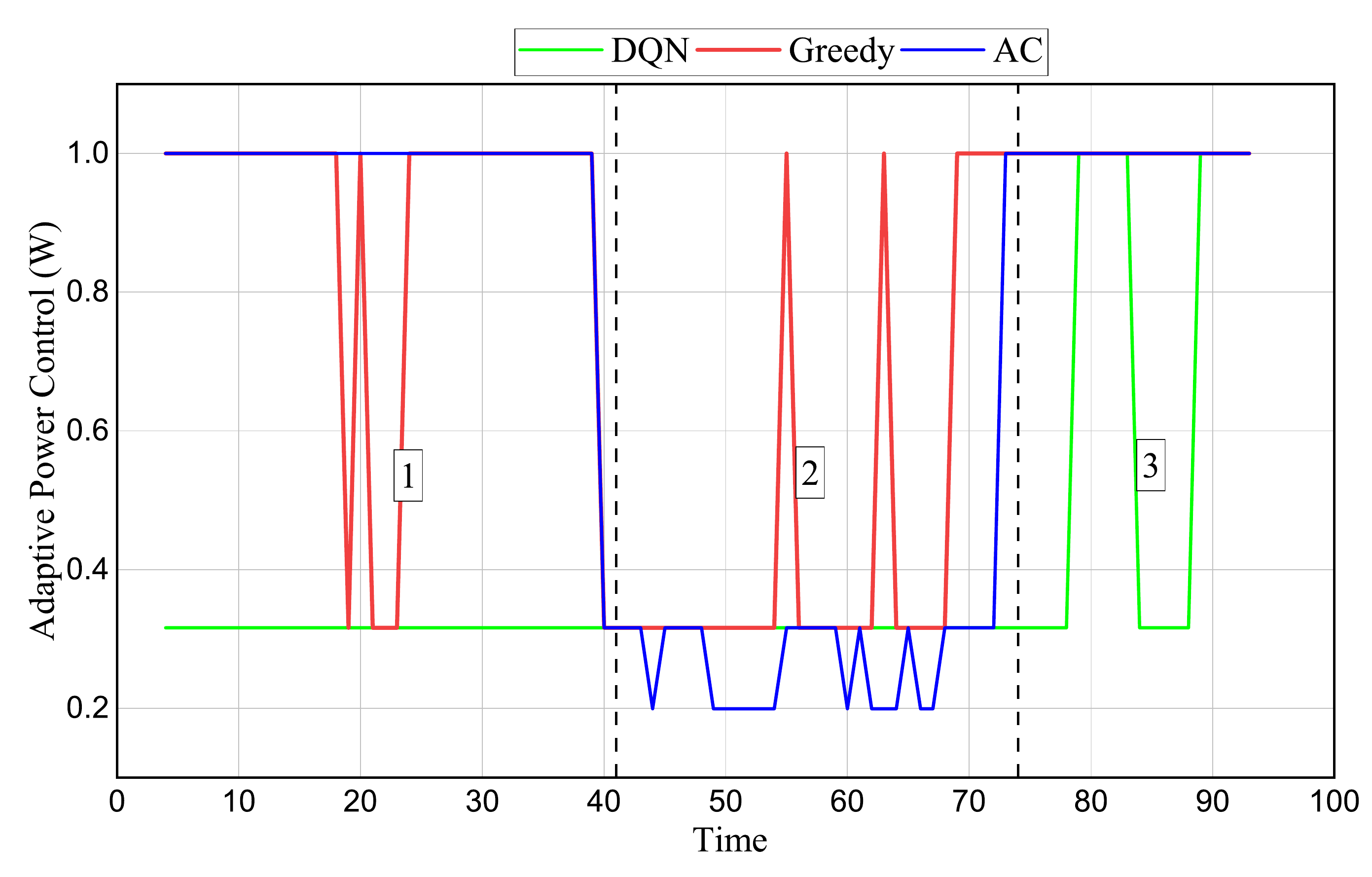}
    \caption{The power control of the UAV-UEs in continuous time slots with different learning algorithms.}
    \label{fig:performance_p}
\end{figure}

\begin{figure}[!t]
    \centering
    \includegraphics[width=3.5 in,height=2.5in]{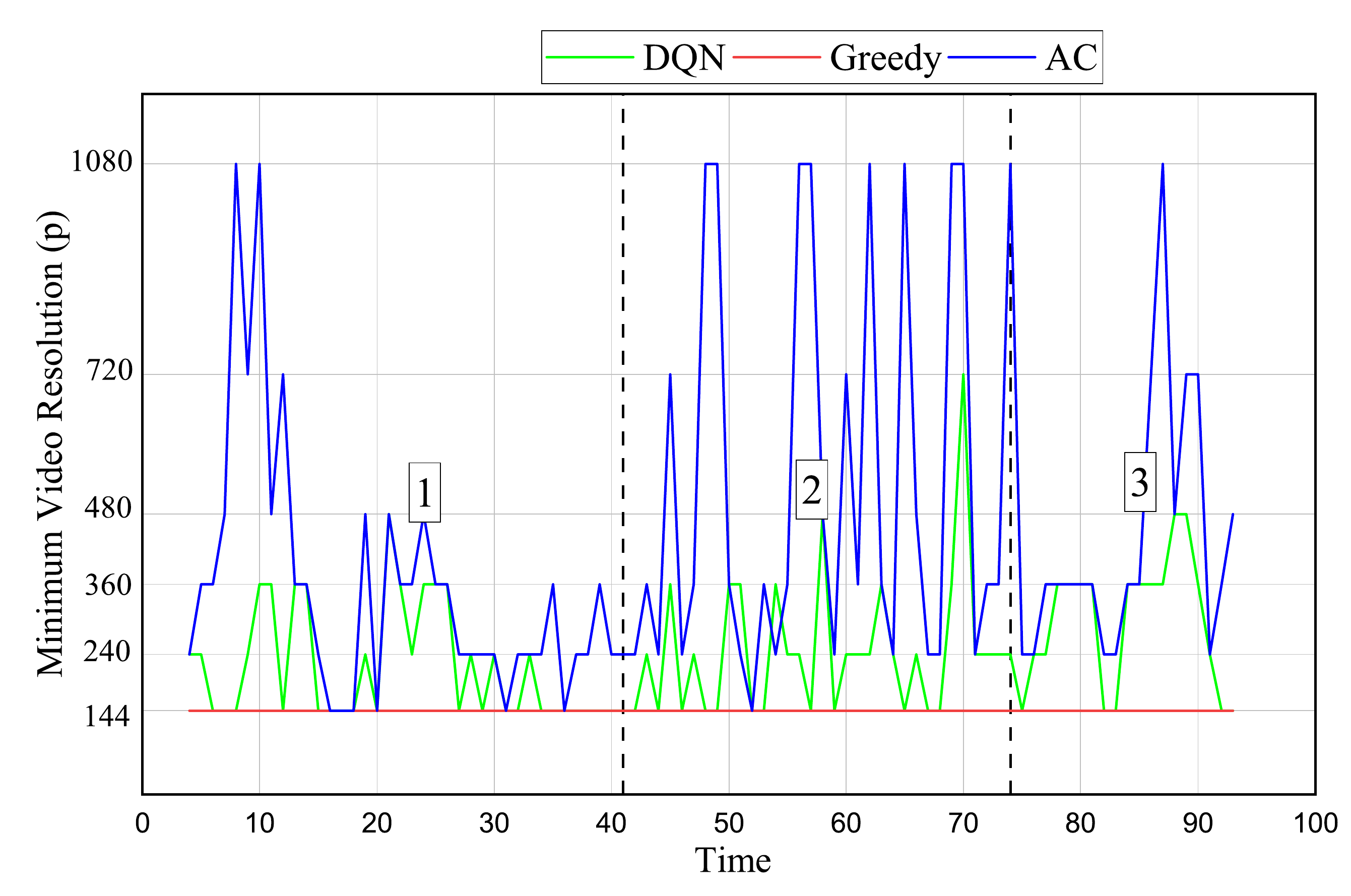}
    \caption{The average adaptive resolution of the UAV-UEs in continuous time slots with different learning algorithms.}
    \label{fig:performance_reso}
\end{figure}
Fig. \ref{fig:reward} plots the average QoE of the UAV-BS with different
video transmission schemes via different learning algorithms in each episode. For simplicity,
``Adaptive Resolution'' represents the scheme with
adaptive resolution, ``AB'' is the scheme with
adaptive resolution and dynamic UAV-BS, and ``ABU''
is the scheme with adaptive resolution, dynamic UAV-BS
and UAV-UEs. It is observed that the average QoE  of
the AC algorithm outperforms all other algorithms,
with it being able to achieve an optimal trade-off between
data rate, bitrate resolution selection, power control, and
positions. From the result, it is observed that with the dynamic
environment and large size of the action, and the AC algorithm is able to select
proper positions of UAVs and video resolution of video frames.
This is mainly due to the experience replay mechanism,
which efficiently utilizes the training samples, and the actor and critic functions are able to smooth the training
distribution over the previous behaviours compared to DQN. In
addition, we can observe that the strategies of selecting optimal
positions for UAVs achieve higher performance compared to
the UAVs with fixed locations. This result emphasizes the
importance of the strategy with mobile UAVs. This is due to the fact that
mobile UAVs can move through the network to reach the
optimal positions that are able to adapt to dynamic fire
scenarios.

Next, we provide more in-depth investigation of the relationship
between the number of UAV request, adaptive video resolution, adaptive power control, and throughput
with different learning algorithms in continuous 100
time slots. The results are also compared among the
three algorithms, namely DQN, AC, and Greedy algorithms. The detailed results show how the control optimization  helps UAVs to
maximize the QoE at each time slot. 

\begin{figure}[!t]
    \centering
    \includegraphics[width=3.5 in,height=2.5in]{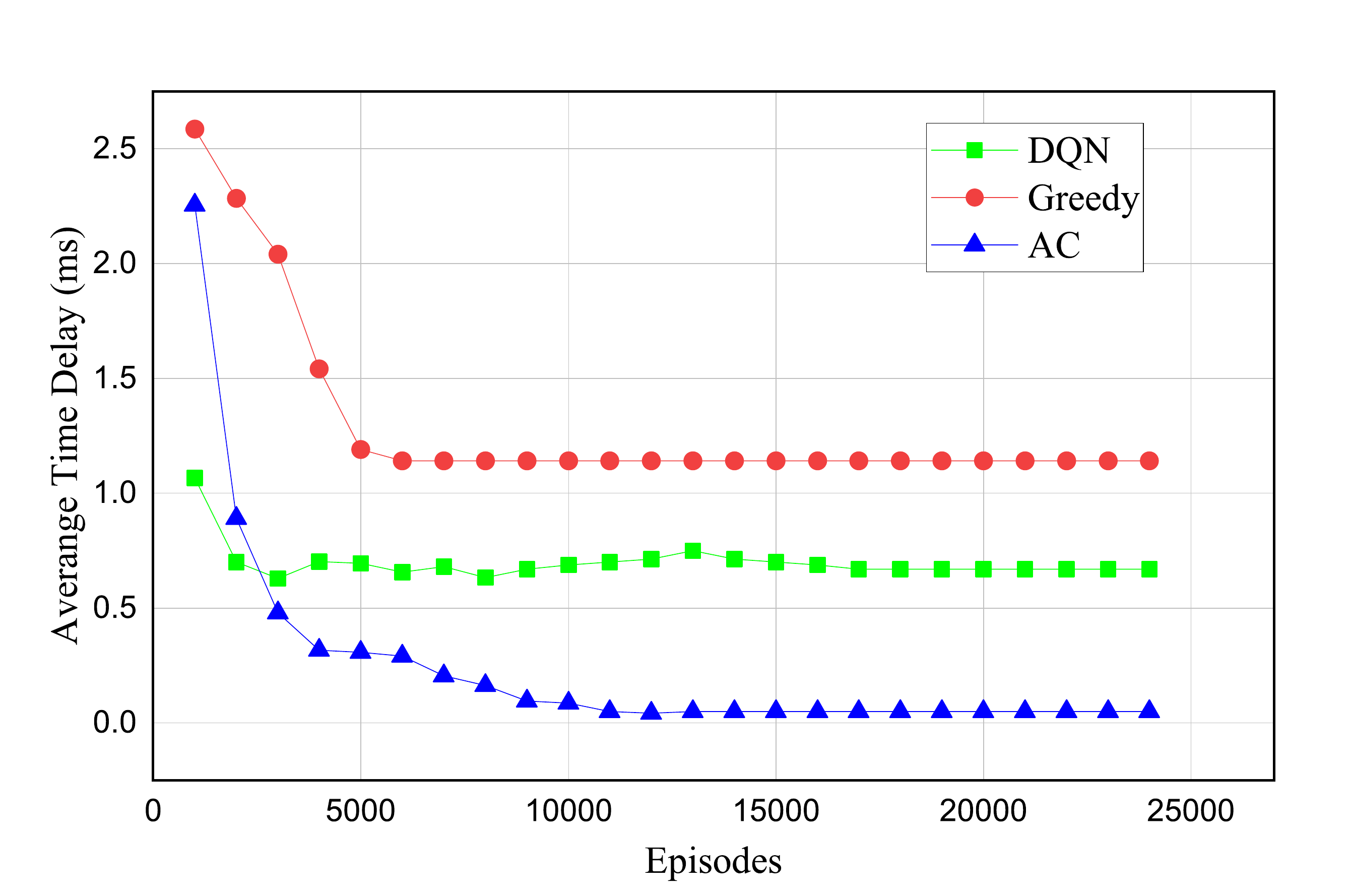}
    \caption{Average latency of video streaming with different learning
algorithms.}
    \label{fig:time}
\end{figure}

Fig. \ref{fig:request} plots the UAV's requests follow the fire arrival distribution, which follow Poisson process distribution with density $\lambda$.
In phase 1, there is a small number of fire arrival which leads to low request of UAV's number. But, when time is increasing, the number od fire arrival is getting higher and leads high number of UAV's request is needed as shown in phase 2 and in phase 3, the request is drop and less UAV's request is demand. 
As the number of requests rapidly changes, we introduce power control to control the transmit power at UAV-UEs to mitigate the interference among UAV-UEs, thus maximizing the achievable rate of each UAV-UE. 

Following the fire arrival requests in Fig \ref{fig:request}, Fig. \ref{fig:performance_p} plots the average power control  over all UAV-UEs in continuous time slots with AC, DQN and Greedy algorithms. The
power control helps mitigating the interference among UAV-UEs. As shown in
phase 1 and phase 3 in Fig. \ref{fig:request}, there is a small number of fire
requests with small number of UAVs to transmit
the data. However, when the number of requests increases, a
large number of UAVs are demanded as shown in phase 2 of Fig. \ref{fig:request}.
As can be seen from Phase 2 of Fig. \ref{fig:performance_p}, the DRL algorithms learn the environment and effectively reduce the transmit  power of each UAV-UE, to reduce the interference from UAV-UEs. We see that
the Greedy algorithm maintains the higher
power, even though high power can provide high received signal, it also causes  high interference at the UAV-BS and failure
in transmission.
\begin{figure}[!t]
    \centering
    \includegraphics[width=3.5 in,height=2.5in]{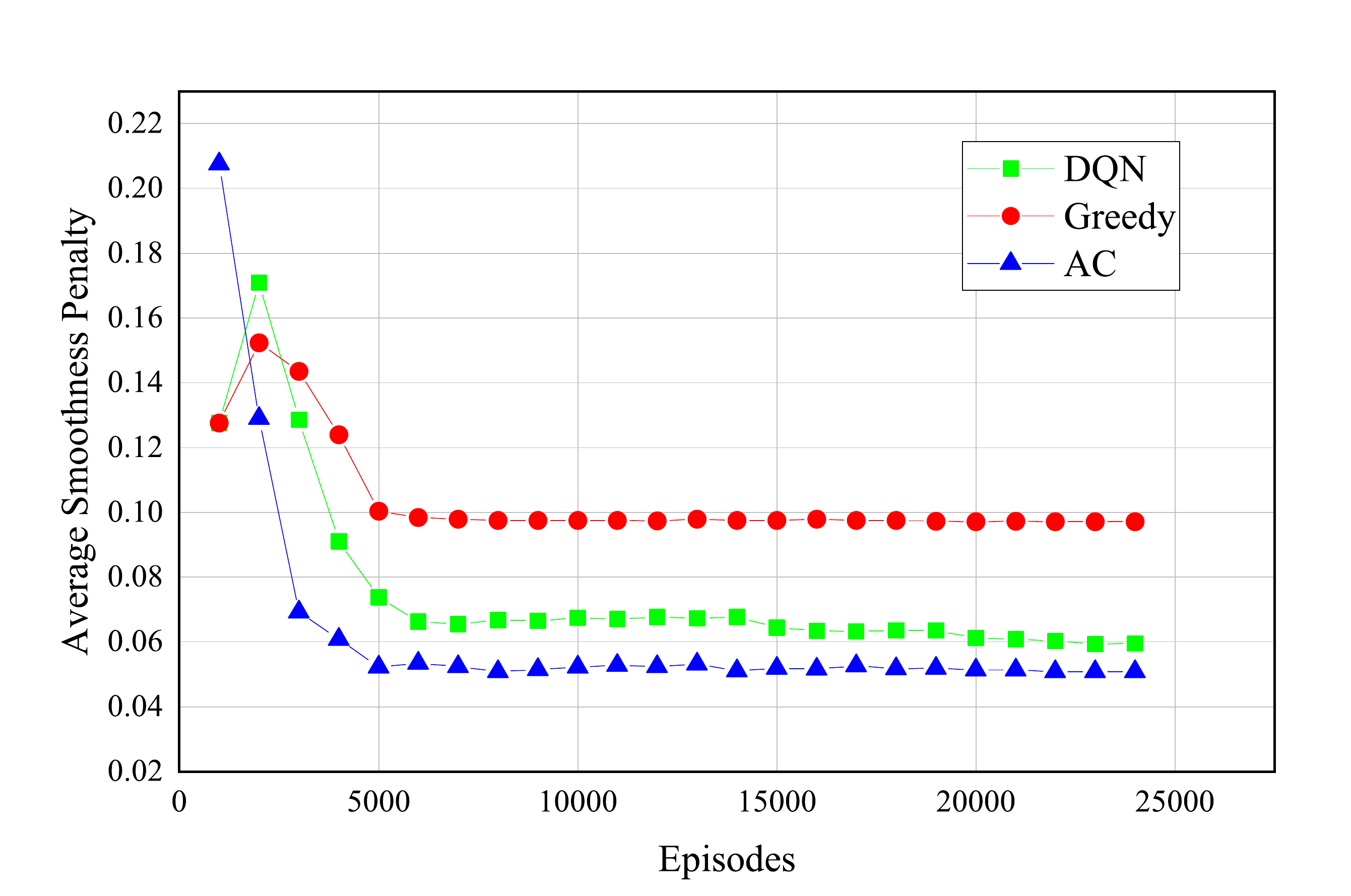}
    \caption{Average smoothness penalty with different learning algorithms.}
    \label{fig:smooth}
\end{figure}

Following the fire arrival requests in Fig. \ref{fig:request}, Fig. \ref{fig:performance_reso} plots the  minimum adaptive resolution over all UAV-UEs in continuous time slots with different learning algorithms. 
It is shown that the minimum video resolution  of the AC algorithm is higher than that of the DQN and the Greedy algorithm in all scenarios.
The AC algorithm is able to maintain an optimal video resolution at each time slot and guarantee high quality and smooth video playback with new request. However,  the Greedy algorithm exploits with a minimum video resolution to maintain high rewards, and it only uses local optimal policy and causes poor performance.
For phase 1 and 3, when  the number of requests is low at the $t$th time slot, the power is high, and the throughput increases, thus, the resolution of video is high.  
However, when the number of request is increasing in phase 2, the AC algorithm is able to maintain a high resolution due to helps of adaptive power, which leads to better QoE for each UAV-UE. This will help to reduce the interference and improve the quality of the video resolution.


In Fig. \ref{fig:time}, we plot the average latency of video streaming with AC, DQN and Greedy  algorithms. It can be seen that the latency performance of the AC algorithm outperforms that of the DQN algorithm. When multiple video streaming exist in the U2U communication, the interference among UAV-UEs occur and cause higher latency. Based on the observed state, the AC algorithm is able to select proper positions and transmission power of the UAV-UEs to mitigate the interference, which further decreases the latency. Thus, the AC algorithm is able to maximize average QoE with the lowest average time latency. However, the Greedy algorithm is unable to exploit the violation of latency constraints and lead to higher latency, which leads to lower QoE.

Fig. \ref{fig:smooth} plots the average smoothness penalty with AC, DQN and Greedy  algorithms. The smoothness penalty demonstrates the average video stability occupancy of UAV-UEs at each episode. 
When the learning algorithm is able to automatically choose the suitable resolution at the $t$th time slots and $(t-1)$th time slot, it will obtain lower smoothness penalty and higher QoE. 
Moreover, the AC algorithm is able to automatically choose the proper action based on actor and critic function which leads to better smoothness  of the AC algorithm compared to  that of the DQN and Greedy algorithms. It is proved that the AC algorithm guarantees the smoothness of video transmission with high QoE. Meanwhile, the Greedy algorithm shows the worst performance because it only makes local optimal selections.

\section{Conclusion} \label{conclusion}
In this paper, we developed a deep reinforcement learning approach for the mobile U2U communication to maximize the Quality of Experience (QoE) of UAV-UEs,  through optimizing the locations for all UAVs, the additive video resolution, and transmission power for UAV-UEs. The dynamic interference problem was handled by utilizing adaptive power control to achieve  a higher achievable rate. Through our developed  Deep Q Network and Actor-Critic methods, the optimal additive video resolution can be  selected to stream real-time video frames, and optimal positions of the UAV-BS and UAV-UEs can be  selected to satisfy the transmission rate requirement. Simulation results demonstrated  the effectiveness of our proposed learning-based schemes  compared to the Greedy algorithm in terms of higher QoE with low latency and high video smoothness.
\bibliographystyle{IEEEtran}
\bibliography{IEEEabrv,main}

\begin{thebibliography}{10}
\providecommand{\url}[1]{#1}
\csname url@samestyle\endcsname
\providecommand{\newblock}{\relax}
\providecommand{\bibinfo}[2]{#2}
\providecommand{\BIBentrySTDinterwordspacing}{\spaceskip=0pt\relax}
\providecommand{\BIBentryALTinterwordstretchfactor}{4}
\providecommand{\BIBentryALTinterwordspacing}{\spaceskip=\fontdimen2\font plus
\BIBentryALTinterwordstretchfactor\fontdimen3\font minus
  \fontdimen4\font\relax}
\providecommand{\BIBforeignlanguage}[2]{{%
\expandafter\ifx\csname l@#1\endcsname\relax
\typeout{** WARNING: IEEEtran.bst: No hyphenation pattern has been}%
\typeout{** loaded for the language `#1'. Using the pattern for}%
\typeout{** the default language instead.}%
\else
\language=\csname l@#1\endcsname
\fi
#2}}
\providecommand{\BIBdecl}{\relax}
\BIBdecl

\bibitem{muller2020forest}
M.~M{\"u}ller, L.~Vil{\`a}-Vilardell, and H.~Vacik, ``Forest fires in the
  alps--state of knowledge, future challenges and options for an integrated
  fire management,'' \emph{EUSALP Action Group}, vol.~8, 2020.

\bibitem{sung2019primo_firefighting}
K.~W. Sung \emph{et~al.}, ``{PriMO-5G:} making firefighting smarter with
  immersive videos through {5G},'' in \emph{Proc. 2019 IEEE 2nd 5G World Forum
  (5GWF)}, Sep. 2019, pp. 280--285.

\bibitem{azari2019u2u}
M.~M. Azari, G.~Geraci, A.~Garcia-Rodriguez, and S.~Pollin, ``Cellular
  {UAV-to-UAV} communications,'' in \emph{Proc. IEEE 30th Annu. Int. Symp.
  Pers. Indoor Mobile Radio Commun. (PIMRC)}, Sep. 2019, pp. 120--127.

\bibitem{zhang2019u2x}
S.~Zhang, H.~Zhang, B.~Di, and L.~Song, ``Cellular {UAV-to-X} communications:
  Design and optimization for multi-{UAV} networks,'' \emph{IEEE Trans.
  Wireless Commun.}, vol.~18, no.~2, pp. 1346--1359, Feb. 2019.

\bibitem{azari2020uav}
M.~M. Azari, G.~Geraci, A.~Garcia-Rodriguez, and S.~Pollin, ``{UAV-to-UAV}
  communications in cellular networks,'' \emph{IEEE Trans. on Wireless
  Commun.}, vol.~19, no.~9, pp. 6130--6144, Jun. 2020.

\bibitem{UAV_disaster}
X.~Liu \emph{et~al.}, ``Transceiver design and multihop {D2D for UAV IoT}
  coverage in disasters,'' \emph{IEEE Internet of Things J.}, vol.~6, no.~2,
  pp. 1803--1815, Apr. 2019.

\bibitem{joshi2020simulation}
A.~Joshi, S.~Dhongdi, S.~Kumar, and K.~Anupama, ``Simulation of multi-uav
  {Ad-Hoc} network for disaster monitoring applications,'' in \emph{2020 Int.
  Conf. on Inf. Network. (ICOIN)}, Jan. 2020, pp. 690--695.

\bibitem{masaracchia2020concept}
A.~Masaracchia \emph{et~al.}, ``The concept of time sharing {NOMA} into
  {UAV-Enabled} communications: An energy-efficient approach,'' in \emph{2020
  4th Int. Conf. on Recent Advances in Signal Processing, Telecommunications \&
  Comput. (SigTelCom)}, Aug. 2020, pp. 61--65.

\bibitem{challita2018deep}
U.~Challita, W.~Saad, and C.~Bettstetter, ``Deep reinforcement learning for
  interference-aware path planning of cellular-connected {UAVs},'' in
  \emph{Proc. 2018 IEEE Int. Commun. Conf. (ICC)}.\hskip 1em plus 0.5em minus
  0.4em\relax IEEE, Jul. 2018, pp. 1--7.

\bibitem{sadi2014minimum}
Y.~Sadi, S.~C. Ergen, and P.~Park, ``Minimum energy data transmission for
  wireless networked control systems,'' \emph{IEEE Trans. on Wireless Commun.},
  vol.~13, no.~4, pp. 2163--2175, Feb. 2014.

\bibitem{zhang2017joint}
S.~Zhang \emph{et~al.}, ``Joint trajectory and power optimization for {UAV}
  relay networks,'' \emph{IEEE Commun. Lett.}, vol.~22, no.~1, pp. 161--164,
  Oct. 2017.

\bibitem{padilla2020flight}
G.~E.~G. Padilla, K.-J. Kim, S.-H. Park, and K.-H. Yu, ``Flight path planning
  of solar-powered {UAV} for sustainable communication relay,'' \emph{IEEE
  Robot. Automat. Lett.}, vol.~5, no.~4, pp. 6772--6779, Aug. 2020.

\bibitem{selim2019outage}
M.~M. Selim \emph{et~al.}, ``On the outage probability and power control of
  {D2D} underlaying {NOMA UAV-assisted} networks,'' \emph{IEEE Access}, vol.~7,
  pp. 16\,525--16\,536, Jan. 2019.

\bibitem{xiao2019bitrate}
X.~Xiao \emph{et~al.}, ``Sensor-augmented neural adaptive bitrate video
  streaming on {UAVs},'' \emph{IEEE Trans. on Multimedia}, pp. 1--12, Oct.
  2019.

\bibitem{govil2020preliminary_fire}
K.~Govil, M.~L. Welch, J.~T. Ball, and C.~R. Pennypacker, ``Preliminary results
  from a wildfire detection system using deep learning on remote camera
  images,'' \emph{Remote Sensing}, vol.~12, no.~1, p. 166, 2020.

\bibitem{nanjiang2018deep}
N.~Jiang, Y.~Deng, A.~Nallanathan, and J.~A. Chambers, ``Reinforcement learning
  for real-time optimization in {NB-IoT} networks,'' \emph{IEEE J. Sel. Areas
  Commun.}, vol.~37, no.~6, pp. 1424--1440, Jun. 2019.

\bibitem{podur2010fire}
J.~J. Podur, D.~L. Martell, and D.~Stanford, ``A compound poisson model for the
  annual area burned by forest fires in the province of ontario,''
  \emph{Environmetrics}, vol.~21, no.~5, pp. 457--469, 2010.

\bibitem{val2018global_fire}
M.~Val~Martin, R.~Kahn, and M.~Tosca, ``A global analysis of wildfire smoke
  injection heights derived from space-based multi-angle imaging,''
  \emph{Remote Sensing}, vol.~10, no.~10, p. 1609, Oct. 2018.

\bibitem{gov.uk_2017}
\BIBentryALTinterwordspacing
``Drones: how to fly them safely and legally,'' Sep 2017. [Online]. Available:
  \url{https://www.gov.uk/government/news/drones-are-you-flying-yours-safely-and-legally}
\BIBentrySTDinterwordspacing

\bibitem{durgin2003space}
G.~D. Durgin, \emph{Space-time wireless channels}.\hskip 1em plus 0.5em minus
  0.4em\relax Prentice Hall Professional, 2003.

\bibitem{goddemeier2015investigation}
N.~Goddemeier and C.~Wietfeld, ``Investigation of air-to-air channel
  characteristics and a {UAV} specific extension to the rice model,'' in
  \emph{2015 IEEE Globecom Workshops (GC Wkshps)}, Dec. 2015, pp. 1--5.

\bibitem{yajnanarayana2018interference}
V.~Yajnanarayana \emph{et~al.}, ``Interference mitigation methods for unmannedd
  aerial vehicles served by cellular networks,'' in \emph{2018 IEEE 5G World
  Forum (5GWF)}, Jul. 2018, pp. 118--122.

\bibitem{3gpp.36.777}
``Study on enhanced lte support for aerial vehicles,'' {3GPP}, TR 36.777, Dec.
  2017, {V15.0.0}.

\bibitem{quality_video}
\BIBentryALTinterwordspacing
``Recommended upload encoding settings - youtube help.'' [Online]. Available:
  \url{https://support.google.com/youtube/answer/1722171?hl=en-G}
\BIBentrySTDinterwordspacing

\bibitem{carballeira2012framework_parallel}
P.~Carballeira, J.~Cabrera, A.~Ortega, F.~Jaureguizar, and N.~Garc{\'\i}a, ``A
  framework for the analysis and optimization of encoding latency for multiview
  video,'' \emph{IEEE J. Sel. Topics Signal Process.}, vol.~6, no.~5, pp.
  583--596, Sept. 2012.

\bibitem{yin2015streaming}
X.~Yin, A.~Jindal, V.~Sekar, and B.~Sinopoli, ``A control-theoretic approach
  for dynamic adaptive video streaming over {HTTP},'' in \emph{Proc. 2015 ACM
  Conf. on Special Interest Group on Data Commun.}, Aug. 2015, p. 325–338.

\bibitem{Mao2017}
H.~Mao, R.~Netravali, and M.~Alizadeh, ``Neural adaptive video streaming with
  pensieve,'' in \emph{Proc. Conf. of the ACM Special Interest Group on Data
  Commun.}, Aug. 2017, pp. 197--210.

\bibitem{mnih2015human_rl}
V.~Mnih \emph{et~al.}, ``Human-level control through deep reinforcement
  learning,'' \emph{Nature}, vol. 518, no. 7540, p. 529, Feb. 2015.

\bibitem{zhang2019qoe}
Z.~Zhang \emph{et~al.}, ``{QoE} aware transcoding for live streaming in
  {SDN-Based Cloud-Aided HetNets}: An actor-critic approach,'' in \emph{Proc.
  2019 IEEE Int. Commun. Conf. Workshops (ICC Workshops)}, May 2019, pp. 1--6.

\bibitem{al2014optimalLAP}
A.~Al-Hourani, S.~Kandeepan, and S.~Lardner, ``Optimal {LAP} altitude for
  maximum coverage,'' \emph{IEEE Commun. Lett.}, vol.~3, no.~6, pp. 569--572,
  Dec. 2014.

\bibitem{urllc_uav}
C.~{She} \emph{et~al.}, ``Ultra-reliable and low-latency communications in
  unmanned aerial vehicle communication systems,'' \emph{IEEE Trans. Commun.},
  vol.~67, no.~5, pp. 3768--3781, May 2019.

\bibitem{al2014modelingPL}
A.~Al-Hourani, S.~Kandeepan, and A.~Jamalipour, ``Modeling air-to-ground path
  loss for low altitude platforms in urban environments,'' in \emph{2014 IEEE
  Global Commun. Conf.}, Dec. 2014, pp. 2898--2904.

\end{thebibliography}
\newpage
\end{document}